# Magnetic skyrmion lattice disclinations in pentagon- and heptagon-shaped FeGe crystals

Thibaud Denneulin* and Rafal E. Dunin-Borkowski
Ernst Ruska-Centre for Microscopy and Spectroscopy with Electrons,
Forschungszentrum Jülich, 52425 Jülich, Germany.

Nikolai S. Kiselev
Peter Grünberg Institute and Institute for Advanced Simulation,
Forschungszentrum Jülich and JARA, 52425 Jülich, Germany.

Vladyslav M. Kuchkin
Department of Physics and Materials Science,
University of Luxembourg, L-1511 Luxembourg.

Magnetic skyrmions in chiral magnets typically arrange into hexagonal lattices, with their structural order influenced by factors such as temperature, external magnetic fields and geometric constraints. While translational defects in skyrmion lattices such as dislocations have been extensively studied, individual angular defects, or *disclinations*, remain largely unexplored. Here, we report on the stabilization of five-fold and seven-fold disclinations in pentagon- and heptagon-shaped FeGe nanocrystals created using focused ion beam milling. The magnetic and elastic structures of the disclinated lattices are investigated using Fresnel imaging and off-axis electron holography in a transmission electron microscope. The results are supported by micromagnetic simulations and analytical models based on linear elasticity theory.

## I. INTRODUCTION

Five-fold symmetry is common in biological systems, for instance, in flowers, starfishes, viruses [1], and protein supermolecules [2]. In contrast, it is much less common in crystals as it is geometrically impossible to create a five-fold periodic lattice. There are a few historical reports of *fiveling* in mineralogy, in macrometer scale crystals of gold and copper [3, 4]. In a fiveling (or pentatwin), five wedge-shaped single crystals separated by twin boundaries are distributed around a common axis and produce a pentagonal shape. In the 1960's, different studies have reported that synthesized metallic nanoparticles can exhibit pentatwins based on Transmission Electron Microscopy (TEM) observations [5–7]. A pentatwin can also be described as a *disclination*, a type of crystalline defect in which rotational symmetry is violated [8, 9]. The term was derived from *disinclination*, which was first introduced in the field of liquid crystals [10]. There are also certain structural relations between pentatwins and quasicrystals discovered later in the 1980's [11]. Due to the missing wedge, a disclinated crystal experiences an elastic deformation which varies circularly around the quintuple junction [12, 13]. In two-dimensional hexagonal systems, 5-fold and 7-fold disclinations can also be created by the removal or the addition of a 60° wedge [14]. 2D disclinations have been described, for instance, in bubble raft crystals on water [15], in topological crystalline insulators [16] and in graphene [17]. In a graphene nanotube, a 5-fold disclination can be observed at the apex, facilitating the closure of the nanotube [18]. More generally, disclinations are a key ingredient of curved geometries [19].

* t.denneulin@fz-juelich.de

*Magnetic skyrmions* are swirling spin textures possessing properties of classical particles with potential applications in the field of spintronics [20]. They were first observed at low temperature in non-centrosymmetric B20 compounds, such as MnSi and FeGe [21, 22], where they are stabilized by the balance between the exchange interaction and the Dzaloshinskii-Moriya interaction [23, 24]. Skyrmions in chiral magnets can form two-dimensional hexagonal lattices in particular conditions of temperature and magnetic fields [21, 22, 25]. Similar to atomic lattices, skyrmion lattices can exhibit crystalline defects such as dislocations and grain boundaries depending on the geometric constraints [26–28]. These defects also play a role in the magnetic phase transitions that occur when increasing the external magnetic field or the temperature [29–33]. This *melting* phenomenon and the associated disordering are described by the Kosterlitz, Thouless, Halperin, Nelson, and Young theory (KTHNY-theory) [34–36]. In the presence of large external magnetic fields, 5-7 dislocations (*i.e.* pairs of 5-membered and 7-membered skyrmion rings) dissociate into individual disclinations during the transition between the *solid* and *liquid* phases, also called the *hexatic* phase [30]. Disclinations in chiral magnets have previously been observed in one-dimensional lattices of the *helical* ground state, which form without the need for external magnetic fields [37, 38]. However, disclinations in the *hexatic* phase are particularly difficult to observe due to the high mobility of the skyrmions in the presence of large external magnetic fields. In this study, we report on the stabilization of 5-fold and 7-fold disclinations in the skyrmion *solid* phase of FeGe in pentagon- and heptagon-shaped crystals observed using Lorentz TEM. As structural defects, disclinations are important for the understanding of the physical properties of skyrmion lattices and may play a role in prospective applications in spintronic devices.

## II. RESULTS

### A. Sample description

Figure 1(a) shows a TEM lamella of FeGe with pentagon patterns of different sizes produced by focused ion beam (FIB) milling following the method described in Supplementary Information 1. Contours of carbon of approximately 100 nm thick around the pentagons were created using ion beam-induced deposition. Figure 1(b,c) shows magnified images of a small pentagon and the FeGe/C interface. The interface shows a FIB-damaged layer of $\sim 20 \pm 5$ nm thick, as indicated by dashed lines. The Fourier transforms shown in Fig. 1(d-g), calculated in different regions of the image, show that the inner part of the pentagon is monocrystalline (d,e), whereas the damaged layer and the carbon layer are amorphous (f,g).

Figure 2(a) shows an image of a pentagon with a side length of $l = 320$ nm and a radius of $r = l/(2\sin(36°)) = 272$ nm, which was investigated using Lorentz TEM at a temperature of 220 K. As shown schematically in Fig. 2(b), the 5-fold disclination state is obtained for a total number of skyrmions

$$S = 1 + \sum_{n=1}^{N} 5n, \tag{1}$$

where $n$ is the skyrmion ring number and $N$ is the total number of rings that the pentagon can contain. It can be expected that $N \approx (r/d - 1)$ where $d$ is the inter-skyrmion distance and it was measured previously that $d = 88$ nm in similar conditions [39]. Therefore, for the pentagon size considered here, the maximum number of skyrmion rings is expected to be $N = 2$ and so the total number of skyrmions in the disclination

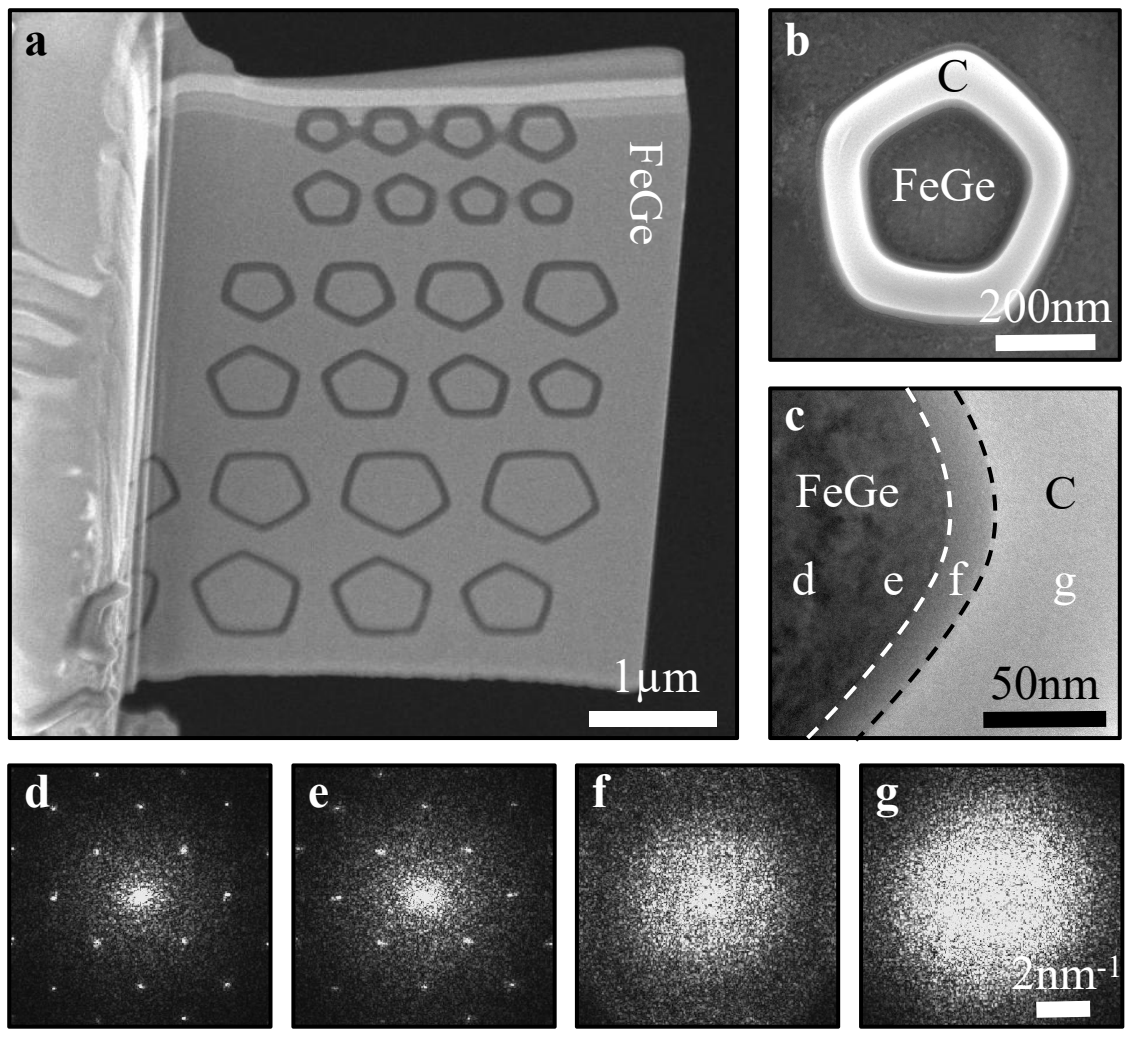

Figure 1. **Sample description.** (a) Representative scanning electron microscopy image of a TEM lamella of FeGe patterned with pentagons of different sizes. (b) Magnified image of a small pentagon. (c) Magnified image of the FeGe/C interface. (d-g) Fourier transforms of different regions indicated in (c).

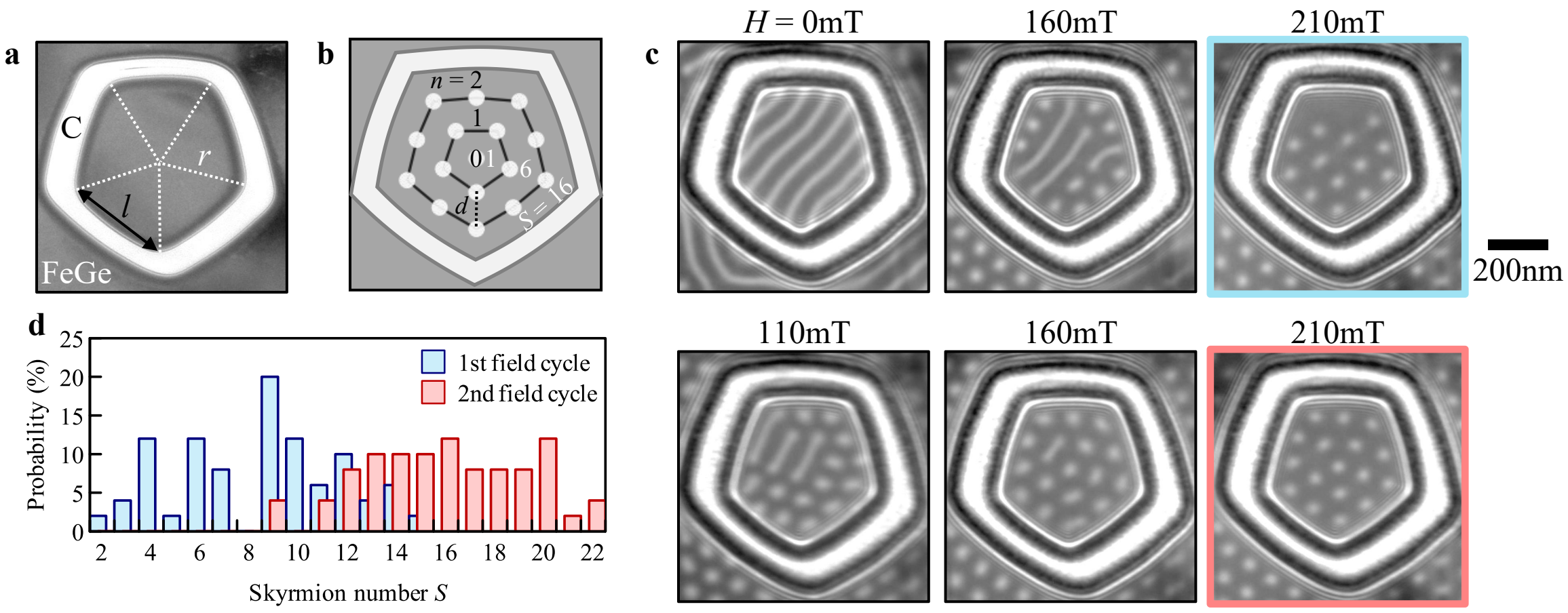

Figure 2. **LTEM of a pentagon-shaped sample with applied magnetic fields.** (a) In-focus TEM image of a pentagon-shaped sample with a side length $l = 320$ nm. (b) Schematic showing a 5-fold disclination with 16 skyrmions. (c) Fresnel images with a 500 µm defocus obtained at 220 K and with different external magnetic fields indicated in the figure. (d) Plot showing the probability to obtain a given skyrmion number $S$ after the first and second field cycles.

state should be $S = 16$.

### B. Magnetic field evolution and skyrmion crystals

Figure 2(c) is a series of Fresnel images that shows the evolution of the magnetic domain structure in the presence of external magnetic fields from the helical ground state at 0 mT to the skyrmion lattice phase at 210 mT (a more detailed field series is shown in Supplementary Information 2). During the first increase of the field, the number of skyrmions is often not large enough to fill the entire pentagon. Here, only 13 skyrmions are obtained at 210 mT. By partially decreasing the magnetic field to 110 mT, additional stripe domains and skyrmions often spontaneously nucleate. Consequently, after a second increase of the field, a larger number of skyrmions is obtained and the lattice can occupy the entire pentagon geometry. Here, the last image obtained at 210 mT shows 17 skyrmions but the number of skyrmions can vary for each new field cycle. Figure 2(d) shows the probability to obtain a given skyrmion number $S$, which was measured by cycling the field 50 times, between 0 and 210 mT in blue and then between 110 and 210 mT in red. The number of skyrmions ranges from 2 to 15 during the first cycle and from 9 to 22 during the second one. The disclination state with 16 skyrmions could not be obtained during the first cycle, but the probability of obtaining it during the second cycle was found to be 12%.

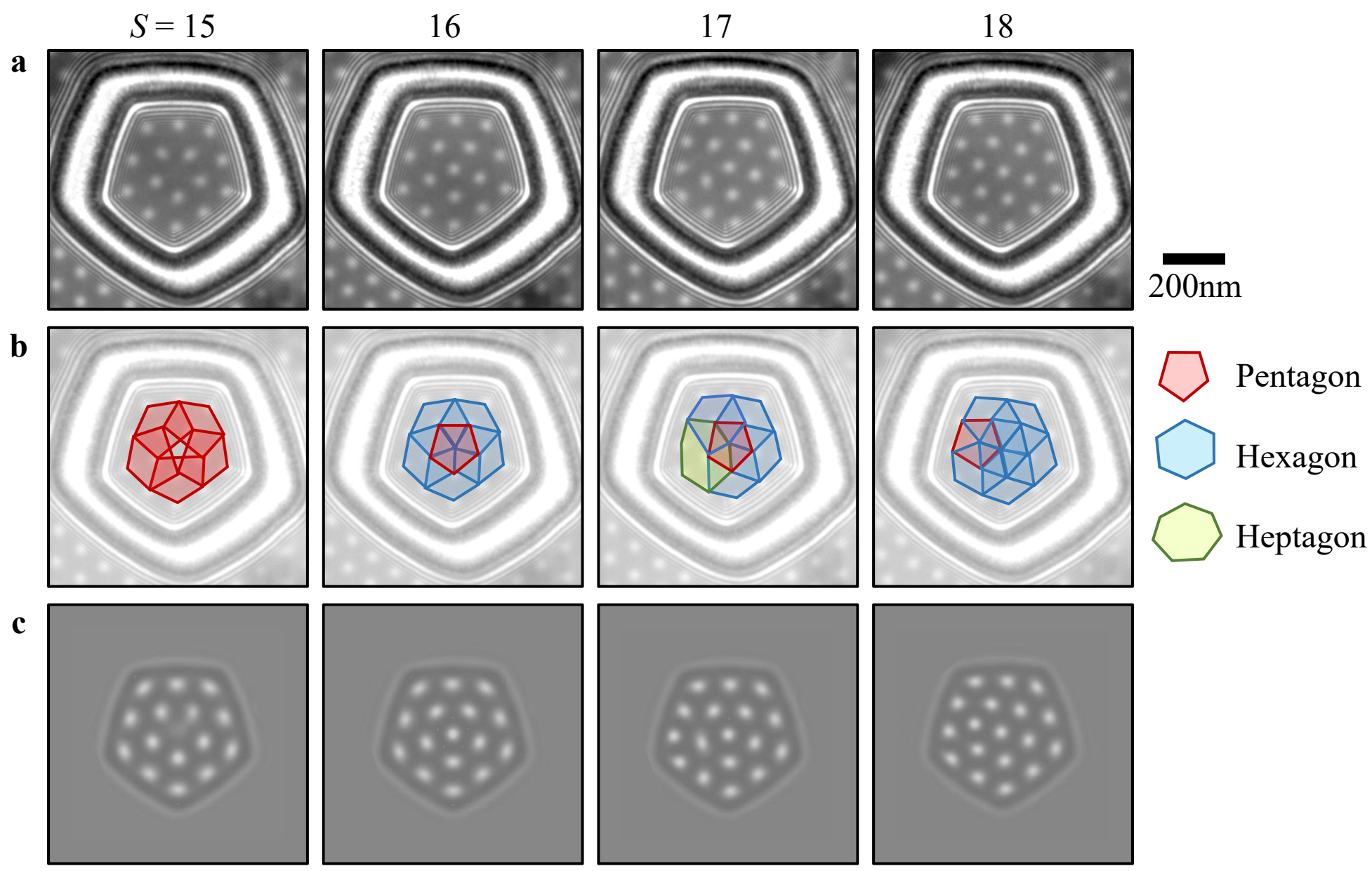

Figure 3. **LTEM of skyrmion lattices in a pentagon-shaped sample.** (a) Fresnel images with a 500 µm defocus obtained at 220 K and with an external magnetic field of 230 mT. The images contain lattices with different skyrmion numbers indicated in the figure. (b) Previous images overlaid with pentagons, hexagons and heptagons. (c) Corresponding Fresnel images calculated from micromagnetic simulations.

Figure 3(a) shows Fresnel images of lattices with different skyrmion numbers ($S = 15$ to 18) obtained by cycling the field. In Fig. 3(b), the lattice was segmented into pentagons, hexagons and heptagons to indicate the number of nearest neighbors. As expected, in the 5-fold disclination state with $S = 16$, the central skyrmion at $n = 0$ has five neighbors whereas the skyrmions of the inner ring $n = 1$ have six neighbors. Interestingly, when $S = 15$, the lattice still shows a 5-fold symmetry but the central skyrmion at $n = 0$ is absent. Fresnel images calculated from micromagnetic simulations in Fig. 3(c) confirm these observations. When $S = 17$, the additional skyrmion with respect to $S = 16$ is placed on the outer ring $n = 2$, which breaks the 5-fold symmetry. In this case, we found it more difficult to identify the exact polygonal structure in the experimental image, probably due to slight irregularities in the shape of the sample. However, the simulation shows that a 5-7 dislocation is expected with the pentagon located in the middle and the heptagon in one of the five corners. When $S = 18$, another skyrmion is added to the inner ring $n = 1$, forming a hexagon in the middle and a pentagon on one side. A similar study carried out on a heptagon-shaped sample is shown in Supplementary Information 3.

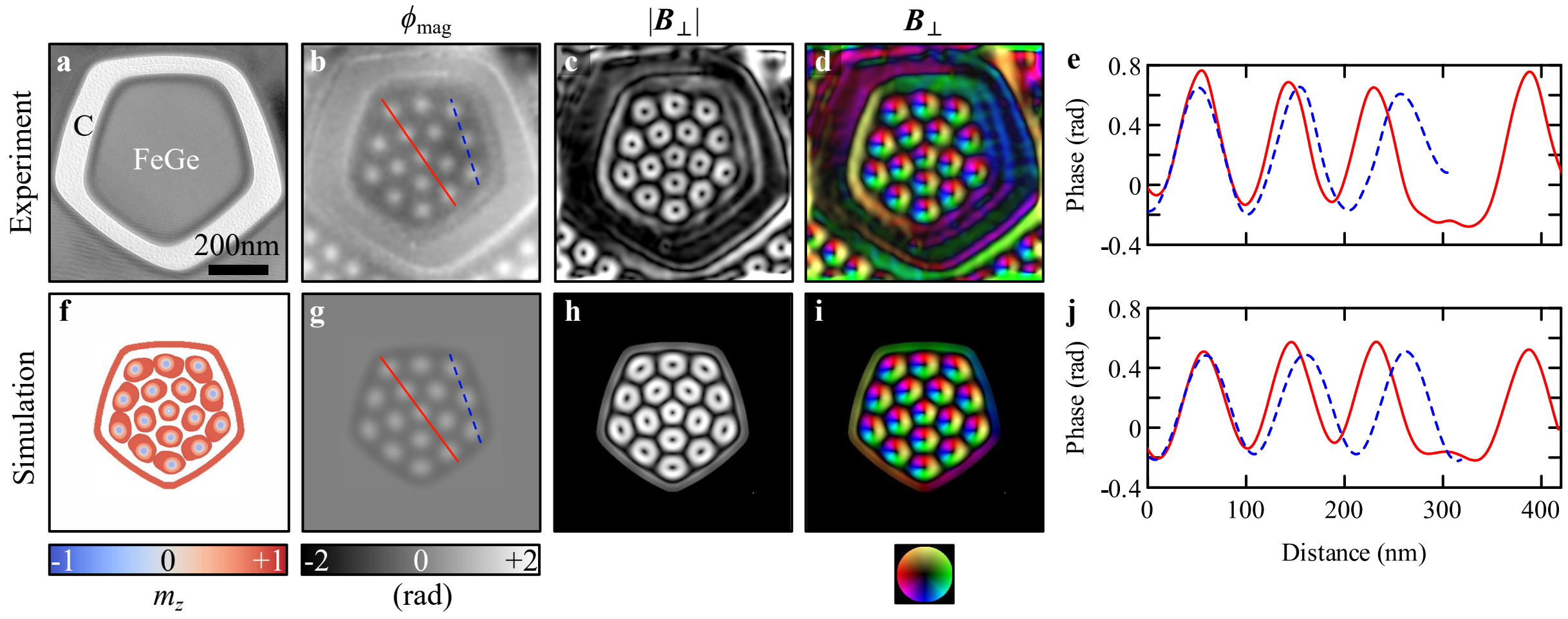

Figure 4. **Electron holography of a 5-fold disclination.** (a) Electron hologram of a pentagon-shaped sample recorded with an external magnetic field of 240 mT. (b) Magnetic phase image $\phi_{mag}$ obtained using off-axis electron holography after subtraction of a phase image recorded at room temperature. (c) Magnitude of the in-plane magnetic induction field $|\boldsymbol{B}_\perp|$. (d) Corresponding color-coded map showing the direction of $\boldsymbol{B}_\perp$. (e) Phase profiles extracted from (b) along the solid and dashed lines. (f) Isosurface representation of the normalized out-plane component $m_z$ of a magnetization model calculated using micromagnetic simulation. (g-j) Corresponding phase image, magnetic field maps and phase profiles.

### C. Magnetic and structural analysis of a 5-fold disclination

To perform a more advanced analysis of the disclination, an in-focus off-axis electron hologram of the pentagon-shaped sample was recorded in Fig. 4(a). Figure 4(b-e) shows the reconstructed *magnetic* phase image $\phi_{mag}$, the magnitude and direction of the in-plane magnetic induction field $\boldsymbol{B}_\perp$, and profiles extracted from the phase image as indicated by the solid and dashed lines. It can be observed in (c) that the central skyrmion has a pentagon shape and is slightly smaller than the other skyrmions. As expected, the magnetic field (d) of the skyrmions rotates around their core, which indicates a Bloch-type texture. The profiles in (e) show that the inter-skyrmion distance along the radial direction (solid red profile) is smaller than the inter-skyrmion distance along the edge direction (dashed blue profile). In a 5-fold disclination, the removal of a $2\pi/6$ wedge induces an elongation of the remaining five wedges along the circular direction. This leads to an increase in the interskyrmion distance along the circular direction. It also leads to an apparent elliptical deformation of the skyrmions, which depends on the applied magnetic field and is more pronounced at low field (see applied field series in Supplementary Information 4).

Figure 4(f-j) shows corresponding micromagnetic simulations of a 5-fold disclination. The simulated images confirm that the central skyrmion is slightly smaller than the surrounding skyrmions, which are elongated along the circular direction around the core. It can be noticed in (f) that this elongation is associated with a twist of the skyrmions along the electron beam direction, which was previously discussed in the context of skyrmion braids [40]. The position of the skyrmions in the simulated phase profiles matches with the experimental profiles and confirm the increase of the inter-skyrmion distance along the circular direction. A similar analysis of a 7-fold disclination is presented in Supplementary Information 5, where inversely, the inter-skyrmion distance is compressed along the circular direction due to the addition of a wedge.

To investigate the structure of the disclination in more details, we identified the center-of-mass coordinates of the 16 skyrmions in the phase image, as shown in Fig. 5(a). The local orientational order parameter $\Psi_6$

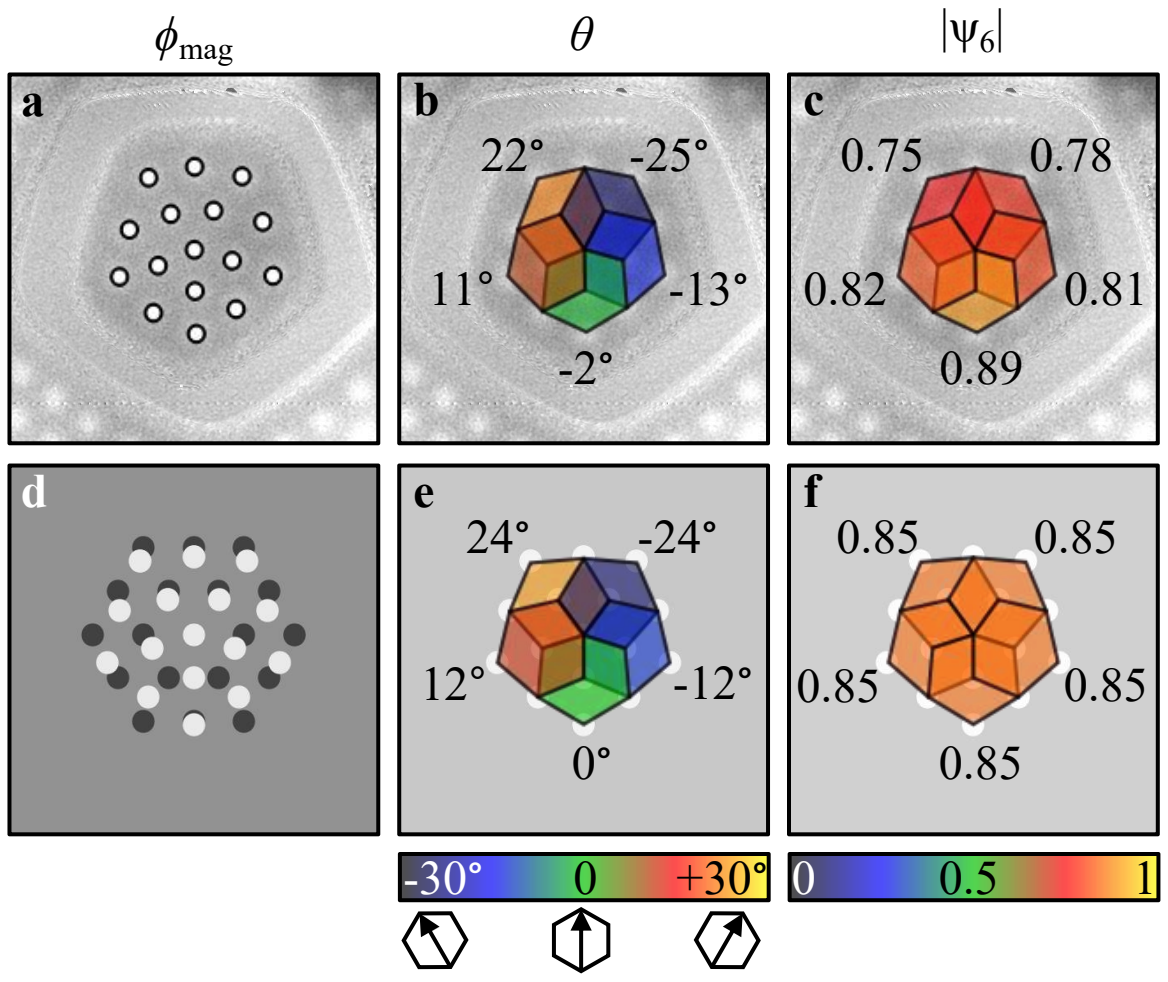

Figure 5. **Structural analysis and linear elastic theory.** (a) Positions of the skyrmions in the magnetic phase image. (b) Orientation $\theta$ and (c) hexagonality $|\Psi_6|$ maps calculated from the positions in (a). (d) 5-fold disclination model calculated using linear elastic theory. (e) Corresponding orientation and (f) hexagonality maps.

of the 5 hexagons was calculated as

$$\Psi_6(r_i) = \frac{1}{6}\sum_{j=1}^{6} \exp(i6\theta_{ij}) \,, \tag{2}$$

with $r_i$ the position of a skyrmion, $\theta_{ij}$ the angle between the line connecting the central skyrmion $i$ and its neighbors $j$ with respect to an arbitrary axis [41]. The parameter $\theta = \arg(\Psi_6)/6$ shown in Fig. 5(b) represents the local orientation of the hexagons. The modulus $|\Psi_6|$ shown in (c) represents the degree of *hexagonality* with $|\Psi_6| = 1$ for a perfect hexagonal structure and $|\Psi_6| < 1$ when it deviates from the hexagonal symmetry. The orientation $\theta$ changes by approximately 12° and in average the hexagonality is $\langle|\Psi_6|\rangle = 0.81$. For comparison, a disclination model was calculated using linear elastic theory. We assumed that the skyrmion lattice can be described as elastically isotropic with a Poisson's ratio $\nu = \frac{1}{3}$ [42]. In isotropic linear elastic theory, the displacement produced by a disclination is given in polar coordinates $(r, \theta)$ by

$$\begin{aligned} u_r &= \frac{\Omega}{4\pi} r \left[ \frac{1-2\nu}{1-\nu} \ln\left(\frac{r}{R}\right) - 1 \right] \,, \\ u_\theta &= \frac{\Omega}{2\pi}\theta \,, \end{aligned} \tag{3}$$

with $\Omega$ the characteristic rotation angle of the disclination ($\Omega = \frac{2\pi}{5}$ in the case of a 5-fold disclination) and $R$ the size of the system [43]. The white dots in Fig. 5(d) show the 5-fold disclination created from a hexagonal lattice (black dots) using Eq. 3. Figure 5(e) and (f) show the corresponding rotation and hexagonality maps calculated from the positions in (d). The rotation changes by $\Delta\theta = 12°$ and the hexagonality is $|\Psi_6| = 0.85$, which is close to the average experimental values. A similar treatment of a 7-fold disclination is also shown in Supplementary Information 5.

### D. Mobility of skyrmions around a disclination

It was observed in Fig. 3 that when $S = 17$ skyrmions, a 5-7 dislocation is created in the lattice. The 7-coordinated skyrmion can be located in any of the five corners of the pentagon, leading to five geometrically and energetically equivalent configurations. Switching between the five configurations can be done by adding an in-plane component to the magnetic field, which can be produced by tilting the sample with respect to the electron beam. Figure 6(a) shows Fresnel images acquired at different tilt angle $\alpha = -3°$, $+3°$ and $0°$. At -3°, the 7-coordinated skyrmion of the dislocation is located in the bottom corner of the pentagon as shown schematically in Fig. 6(b). At +3°, it is located in the bottom-right corner. Interestingly, at an intermediate angle of 0°, the skyrmions along the bottom right edge wobble between the two states leading to a diffuse appearance. In Fig. 6(c), the positions of the skyrmions measured at +3° and -3° are indicated by blue and red circles to show their relative displacements. Between the two tilted states, there is a displacement of the skyrmions located near the bottom-right edge along the circular direction, which explains their diffuse appearance at 0°. Another interesting case of wobbling in a larger pentagon with three rings is described in Supplementary Information 7.

To understand this phenomenon, we performed micromagnetic simulations. The external field was applied with a tilt of 5° with respect to the normal and the in-plane component is oriented perpendicular to one of the pentagon sides, closely matching the experimental conditions. Fig. 6(d) shows the simulated Lorentz images of the lowest energy states at $-5°$ and $+5°$. The 0° image was obtained by averaging the two previous configurations. The minimum energy paths shown in Fig. 6(e) illustrate that, at 0° tilt, the two configurations are almost degenerate, whereas at $\pm5°$ one state becomes significantly lower in energy than the other, depending on the sign of the tilt angle. Furthermore, the energy barrier separating the states increases strongly under tilt, compared to the zero-tilt case. This explains why sequential switching between states, leading to diffuse contrast, occurs only at zero tilt, where the barrier is minimal.

## III. DISCUSSION

Pentagons and heptagon-shaped geometries have been created by focused ion beam in order to stabilize 5-fold and 7-fold disclinations. The disclination state corresponds to a specific number of skyrmions which depends on the size of the sample and the number of skyrmion rings it can contain. The number of skyrmions produced when varying the external magnetic field cannot be controlled but there is a small probability to obtain the disclination when performing multiple cycles (12% for $N = 2$). Experimentally, we were able to obtain disclinations with up to $N = 3$ skyrmion rings (see Supplementary Information 6). However, the probability is particularly low for disclinations of such size.

The structural analysis of the 5-fold and 7-fold disclinations shows respectively, an increase and a decrease of the inter-skyrmion distance along the circular direction, due to the removal or the addition of a wedge. Orientational and hexagonality maps derived from experimental images were found to be in agreement with those calculated from linear elastic theory. The shape of the skyrmions at low field is also modified by the disclination strain. In the 5-fold case, the central skyrmion shows a pentagonal shape and is smaller than the surroundings due to the compressive strain towards the center of the disclination. The surrounding skyrmions show an elliptical deformation along the circular direction around the core, which can be associated with a twist along the thickness of the sample according to simulations. In the case of a 7-fold disclination, the central skyrmion is slightly larger due to the tensile strain and there is an elongation of the surrounding skyrmions along the radial direction.

Finally, the mobility of skyrmions around a disclination was investigated. When the rotation symmetry of the outer skyrmion ring is broken due to the presence of dislocations, there are energetically equivalent configurations. Switching between the different configurations can be done by tilting the sample a few degrees with respect to the out-of-plane field in order to introduce a small in-plane component. When an intermediate angle is reached, the skyrmion lattice can wobble between different configurations leading to a smeared appearance of the skyrmions in the images. This wobbling effect is enabled by the thermal energy at the temperature of the experiment (220 K) that allows the system to overcome the small barrier between the two states. This phenomenon is expected to vanish at low temperature as the system should remain trapped in a local minimum.

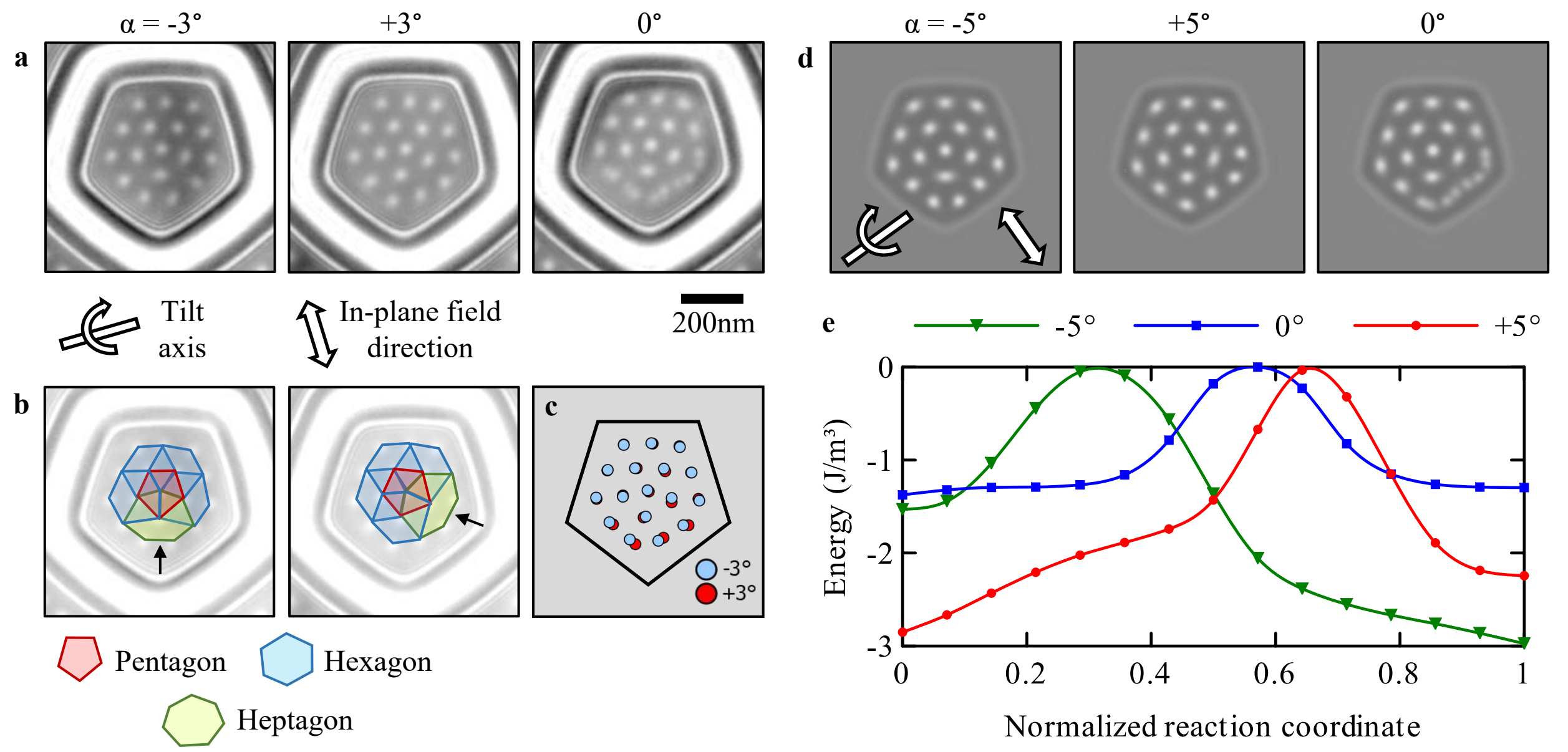

Figure 6. **Skyrmions mobility around a disclination.** (a) Fresnel images of a skyrmion lattice with $S = 17$ obtained in the presence of an external magnetic field of 230 mT at different tilt angles $\alpha = -3°$, $+3°$ and $0°$. (b) Images obtained at $-3°$ and $+3°$ overlaid with pentagons, hexagons and heptagons. (c) Positions of the skyrmions in the two tilted images represented by blue and red dots. (d) Lorentz images calculated from micromagnetic simulations showing the lowest energy configuration at $\alpha = -5°$ and $+5°$ tilt angles. The image at $0°$ is the sum of the previous images. (e) Energy paths for the transition between the two lattice configurations calculated for the three different tilt angles $\pm 5°$ and $0°$.

## IV. CONCLUSION

FeGe single crystals with pentagon and heptagon shape were fabricated by focused ion beam. This allowed us to stabilize 5-fold and 7-fold disclinations in the skyrmion lattice solid phase with relatively low external magnetic fields. The disclinated lattices observed using Lorentz TEM and off-axis electron holography were in agreement with micromagnetic simulations and their structure agreed with linear elastic theory. It was also shown that skyrmion rings around a disclination can wobble between energetically equivalent configurations when dislocations are present. These results demonstrate that geometric confinement is an effective strategy for engineering defects in skyrmion lattices of chiral magnets. Future work could explore the controlled creation/annihilation of individual disclinations using different stimuli such as light, current pulses or mechanical strain.

## ACKNOWLEDGMENTS

This work was supported by the European Union's Horizon 2020 Research and Innovation Programme under grant agreement number 856538 (project "3D MAGIC"). V.M.K. acknowledges the financial support from the European Union's Horizon Europe research and innovation programme under the Marie Skłodowska-Curie grant agreement No. 101203692 (QUANTHOPF).

## AUTHOR CONTRIBUTIONS

T.D. conceived the study, performed the LTEM experiments and wrote the manuscript. N.S.K. performed micromagnetic simulations. V.M.K. performed calculations of energy paths. R.E.D-B. secured funding for the study.

## DATA AVAILABILITY

The data supporting this study will be made openly available upon acceptance of the article at Zenodo: https://doi.org/10.5281/zenodo.18111798. A preview of the data is available at `https://zenodo.org/records/18111798?preview=1&token=eyJhbGciOiJIUzUxMiJ9.eyJpZCI6IjIxYWVmMWRkLTM5ZDgtNDkzNy04MzdhLWFmYzdhNW` `3IXiDx2SJXs6RICzVwpyuGo2swWW5aYRFwlA-huT3k_ZMyxLgdIPzIl_ZKRs1BmV1WdL4aDJosmQXqDSJxe5bg`

## ADDITIONAL INFORMATION

The authors declare no competing interests.

### Appendix A: Sample preparation and Lorentz TEM

TEM lamellae were prepared from a bulk crystal of FeGe using 30 kV focused $Ga^+$ ion beam sputtering in a Thermo Fisher Scientific (TFS) scanning electron microscope (FIB-SEM) Helios dual-beam platform. The samples were patterned following the protocol described in Supplementary Information 1. The final thickness of the lamellae was in the 100 to 150 nm range.

TEM was carried out using two TFS Titan microscopes named *Titan T* [44] and *Titan Holo* [45], both equipped with a Schottky field emission gun operated at 300 kV, a CEOS image aberration corrector, a post-specimen electron biprism and a field free Lorentz mode. *Titan T* is equipped with a 2k×2k Gatan CCD Ultrascan 1000P and *Titan Holo* is equipped with a 4k×4k Gatan K2-IS direct electron detector. The objective lens was used to apply magnetic fields perpendicular to the sample which were precalibrated using a Hall probe. The temperature of the sample was controlled using a Gatan liquid-nitrogen-cooled specimen holder (model 636) and temperature controller (model 1905).

When electrons pass through a magnetic induction field $B_\perp$ that is oriented perpendicular to their trajectory, they are deflected by the Lorentz force. To a first approximation, the deflection angle is given by the expression $\theta = \frac{e\lambda}{h} B_\perp t$, where $e$ is the electron charge, $\lambda$ is the electron wavelength, $t$ is the specimen thickness and $h$ is Planck's constant [46]. When Fresnel images are recorded in a defocused plane, the deflection of the electron beam leads to variations contrast at the position of magnetic domain walls, whose width and amplitude depend on the defocus.

In this manuscript, the images were recorded at a temperature of 220 K with a defocus of 500 μm.

## Appendix B: Off-axis electron holography

In off-axis electron holography, a post-specimen electron biprism is used to overlap a wave traveling through the sample (object wave) with a wave traveling in vacuum (reference wave). The condenser stigmators are used to form an elliptical illumination to improve the coherence in the direction perpendicular to the biprism. We consider a reference wave $\psi_1 = A_1 \exp(i\phi_1)$ traveling in a uniform region (vacuum for instance) and an object wave $\psi_2 = A_2(\vec{r}) \exp(i\phi_2(\vec{r}))$ passing through a sample, where $A_1, A_2(\vec{r})$ and $\phi_1, \phi_2(\vec{r})$ are their respective amplitudes and phases. The intensity distribution of an off-axis electron hologram can be expressed

$$I_{holo}(\vec{r}) = A_1^2 + A_2^2(\vec{r}) + I_{\text{inel}}(\vec{r}) + 2\mu A_1 A_2(\vec{r}) \cos(\Delta\phi(\vec{r}) + 2\pi \vec{q_c}.\vec{r}) \ , \tag{B1}$$

where $I_{\text{inel}}(\vec{r})$ is the inelastic background, $\mu$ is the fringe contrast, $\Delta\phi(\vec{r}) = \phi_1 - \phi_2(\vec{r})$ is the phase change of the object wave with respect to the reference wave and $\vec{q_c}$ is the carrier frequency [47]. The Fourier transform of the hologram can be described as

$$\begin{aligned} \text{FT}\{I_{holo}(\vec{r})\} &= \text{FT}\{A_1^2 + A_2^2(\vec{r}) + I_{\text{inel}}(\vec{r})\} & \textit{centerband} \\ &+\mu \text{FT}\{A_1 A_2(\vec{r}) \exp(i\Delta\phi(\vec{r}))\} \otimes \delta(\vec{q} + \vec{q_c}) & \textit{sideband } 1 \\ &+\mu \text{FT}\{A_1 A_2(\vec{r}) \exp(i\Delta\phi(\vec{r}))\} \otimes \delta(\vec{q} - \vec{q_c}) & \textit{sideband } 2 \ . \end{aligned} \tag{B2}$$

An aperture is applied to a side-band, which is then shifted to the center of Fourier space. Using an inverse Fourier transform, the following images are reconstructed

$$\begin{aligned} C_{\text{rec}}(\vec{r}) &= \mu A_1 A_2(\vec{r}) \exp(i\Delta\phi(\vec{r})) & & , \\ A_{\text{rec}}(\vec{r}) &= \mu A_1 A_2(\vec{r}) &= \sqrt{Re^2 + Im^2} & , \\ \phi_{\text{rec}}(\vec{r}) &= \Delta\phi(\vec{r}) &= \arctan(Im/Re) & , \end{aligned} \tag{B3}$$

where $C_{\text{rec}}(\vec{r})$ is a complex image, $A_{\text{rec}}(\vec{r})$ and $\phi_{\text{rec}}(\vec{r})$ are the corresponding amplitude and phase images.

The phase change $\Delta\phi$ has an electrostatic $\Delta\phi_{elec}$ and a magnetic $\Delta\phi_{mag}$ component according to the expression (in one dimension) [48]

$$\begin{aligned} \Delta\phi(x) &= \Delta\phi_{elec} + \Delta\phi_{mag} \ , \\ \Delta\phi(x) &= C_E \int V_0(x,z)dz - \tfrac{e}{\hbar} \iint B_\perp(x,z)dxdz \ , \end{aligned} \tag{B4}$$

where $z$ is the direction of the incident electron beam, $x$ is a direction perpendicular to $z$, $C_E$ is an interaction constant that depends on the electron energy, $V_0$ is the mean inner potential (MIP) of the specimen, $B_\perp$ is the component of the magnetic induction field that is perpendicular to both $x$ and $z$, $e$ is the electron charge and $\hbar$ is the reduced Planck's constant. Here, the phase variations related to the thickness and the electrostatic component have been removed by subtracting a phase image recorded above the Curie temperature to the phase image obtained at low temperature. The magnetic phase term can then be simplified to

$$\Delta\phi_{mag}(x) = -\frac{et}{\hbar} \int B_\perp(x)dx \ , \tag{B5}$$

where $t$ is the specimen thickness. The derivative of the magnetic phase is proportional to the magnetic induction field according to the expression

$$\frac{d(\Delta\phi_{mag}(x))}{dx} = -\frac{e}{\hbar} B_\perp(x) \ . \tag{B6}$$

Phase images and color-coded magnetic induction maps were reconstructed using the Digital Micrograph software (Gatan) and the Holoworks plugin [49].

To map the local orientation parameter, the subpixel coordinates of the skyrmions' center of mass in the electron phase image were determined after applying a Gaussian filter to the image. Nearest neighbors were then determined manually and the angles $\theta_{ij}$ were calculated from the coordinates of the skyrmions using $atan2$ functions. The complex value of $\Psi_6$, the orientation $\theta = \arg(\Psi_6)/6$ and the hexagonality $|\Psi_6|$ were then calculated for each skyrmion.

### Appendix C: Micromagnetic simulations

We employ the standard micromagnetic model for isotropic chiral magnets, which includes four energy contributions: the Heisenberg exchange, the Dzyaloshinskii–Moriya interaction (DMI), the Zeeman energy from the external magnetic field, and the energy of demagnetizing fields arising from the sample itself. The corresponding energy functional can be written as follows [50]:

$$\mathcal{E} = \int_{V_\mathrm{m}} d\mathbf{r} \left( \mathcal{A} \sum_{i=x,y,z} |\nabla m_i|^2 + \mathcal{D}\mathbf{m}\cdot(\nabla\times\mathbf{m}) - M_\mathrm{s}\mathbf{m}\cdot(\mathbf{B}_\mathrm{ext} + \nabla\times\mathbf{A}_\mathrm{d}) \right) + \frac{1}{2\mu_0}\int_{\mathbb{R}^3} d\mathbf{r} \sum_{i=x,y,z} |\nabla A_{\mathrm{d},i}|^2, \quad \text{(C1)}$$

where $V_\mathrm{m}$ is volume the magnetic sample, $\mathbf{m}(\mathbf{r}) = \mathbf{M}(\mathbf{r})/M_\mathrm{s}$ denotes the magnetization vector field, $M_\mathrm{s}$ is the saturation magnetization, $\mu_0$ is the vacuum permeability, $\mathcal{A}$ and $\mathcal{D}$ are the Heisenberg exchange constant and DMI constant, respectively. $\mathbf{A}_\mathrm{d}(\mathbf{r})$ is the magnetic vector potential induced by the sample magnetization, and $\mathbf{B}_\mathrm{ext}$ is the external magnetic field.

We assume the material parameter for FeGe [40, 50–52]: $\mathcal{A}$ = 4.75 pJ/m, $\mathcal{D}$ = 0.853 mJ/m$^2$, $\mathbf{M}_\mathrm{s}$ = 384 kA/m. The above parameters are optimized for the temperature of 95 K. The experiments presented in this work are performed at the elevated temperature of 220 K. To account for this aspect, we use a reduced value of the saturation magnetization, $M_\mathrm{s}$ = 120 kA/m, which yields quantitative agreement with experimental observations, as shown in Fig. 4 and Supplementary Fig. S5. Simulations were performed on a regular mesh of $256\times256\times64$ cuboids. The cuboid dimensions were chosen to reproduce the geometry of the experimental samples. Specifically, for the pentagon we set the cuboid size to $2.014 \times 2.014 \times 2.12$ nm$^3$, while for the heptagon-shaped sample shown in Supplementary Figs. S3e and S5f–j we used $2.15 \times 2.15 \times 2.12$ nm$^3$. To account for uncertainty in the exact sample size, we varied the simulated dimensions within a range of $\pm10\%$. For the pentagon [Figs. 3(c), 4(f–j), and 6(d)], the best agreement was obtained for a simulated sample 5% smaller than the measured size, whereas for the heptagon the best fit corresponded to the experimentally determined size.

We simulate the FIB-damaged layer by setting $\mathcal{D}$ to zero in a thin layer of cuboids on all edges of the sample [53]. We find the best agreement between experimental and theoretical images for the thickness of the FIB-damaged layer of 24 nm. This corresponds to a layer of approximately 12 cuboids from each side of the sample. For better agreement with the experiment, we modified the sharp linear edges of the pentagon and heptagon to slightly curved ones. More precisely, the sides of the pentagon are approximated by arcs of a circle passing through adjacent vertices of the polyhedron. The radius of this circle is three times larger than the radius of the circumscribed circle of the regular polyhedron. In the case of a pentagon and a heptagon, the centers of the circles are positioned on opposite sides with respect to the polyhedron's side, exhibiting positive and negative curvature, respectively. The micromagnetic simulations were performed using the Excalibur [54] code, which enables efficient energy minimization and *in situ* calculation of Lorentz TEM images according to the method described in the following section. Additionally, we verified our results using the open-source software Mumax [55].

## Appendix D: Simulation of magnetic phase and Fresnel images

Using the phase object approximation, the wave function of an electron beam can be written as follows [56]:

$$\Psi_0(x, y) \propto \exp(i\phi(x, y)) , \tag{D1}$$

where $\varphi(\mathrm{x}, \mathrm{y})$ is the magnetic contribution to the phase shift [47]

$$\phi_{mag}(x, y) = -\frac{e}{\hbar}\int_{-\infty}^{+\infty} A_z dz , \tag{D2}$$

with $e$ an elementary charge and $h$ Planck's constant. The $z$ component $A_z$ of the magnetic vector potential was obtained from micromagnetic simulations. Neglecting aberrations other than defocus, the wave function at the detector plane can be written in the form

$$\Psi_{\Delta z}(x, y) \propto \iint \Psi_0(x', y')K(x - x', y - y')dx'dy' , \tag{D3}$$

where the kernel is given by the expression

$$K(\varepsilon, \eta) = \exp\left(\frac{i\pi}{\lambda\Delta z}(\xi^2 + \eta^2)\right) , \tag{D4}$$

where $\lambda$ is the electron wavelength and $\Delta z$ is the defocus. The Fresnel image intensity is then calculated following

$$I(x, y) \propto |\Psi_{\Delta z}(x, y)|^2 .$$

For more details, see Ref. [40].

## Appendix E: Minimum energy path calculation

To study the transition mechanism between two states corresponding to different packings of 17 skyrmions in the pentagon-shaped sample, we calculated the minimum energy path (MEP) between these configurations using the geodesic nudged elastic band (GNEB) method [57, 58]. The GNEB method yields a sequence of transient states that form a discrete MEP. These transient states, representing snapshots of the system along the transition, are referred to as images. The number of images is a free parameter, chosen to balance the continuity of the energy path with computational efficiency. The method is iterative, starting from an initial guess for the path that is usually constructed by linear interpolation between the initial and final spin configurations. At each iteration, we applied the velocity projection optimization algorithm. According to the GNEB method, all neighboring images must remain equidistant in the geodesic spin space. This constraint is enforced by effective forces, which combine the spin effective fields with an additional spring force proportional to the Euclidean distance between images. Besides the transient states, the GNEB method also identifies the saddle-point configuration along the MEP. From the saddle-point energy, one can estimate the energy barrier that must be overcome for the transition between states. Assuming that the height of the energy barrier is comparable to $k_{\mathrm{B}}T$, the GNEB results can be directly compared with finite-temperature simulations of the stochastic LLG equation [59]. We implemented the GNEB method in our extension [60, 61] of the Mumax software [55] and used it for the MEP calculations shown in Fig. 6(e). For consistency with the micromagnetic simulations, we used the same material parameters: $\mathcal{A} = 4.75$ pJ/m, $\mathcal{D} = 0.853$ mJ/m$^2$, and $M_{\mathrm{sat}} = 120$ kA/m. The shape and size of the simulated domain are identical to those in micromagnetic simulations performed with Excalibur: the mesh density is set to $256 \times 256 \times 64$ cuboids, and each cuboid size is $2.014 \times 2.014 \times 2.12$ nm$^3$. We further assumed a FIB-damaged layer of thickness 24 nm around all sample edges, approximated by setting $\mathcal{D} = 0$ in the corresponding cuboids. For details, see the script provided in Supplementary Materials and initial states are available at https://doi.org/10.5281/zenodo.17234039. Note that this script is compatible only with the extended version of Mumax [60].

---

# Supplementary Information for Magnetic skyrmion lattice disclinations in pentagon- and heptagon-shaped FeGe crystals

Thibaud Denneulin[1], Nikolai S. Kiselev[2], Vladyslav M. Kuchkin[3], Rafal E. Dunin-Borkowski[1]

[1] *Ernst Ruska-Centre for Microscopy and Spectroscopy with Electrons, Forschungszentrum Jülich, 52425 Jülich, Germany.*

[2] *Peter Grünberg Institute and Institute for Advanced Simulation, Forschungszentrum Jülich and JARA, 52425 Jülich, Germany.*

[3] *Department of Physics and Materials Science, University of Luxembourg, L-1511 Luxembourg.*

## Supplementary Information 1: Sample preparation of pentagon and heptagon geometries using focused ion beam

Figure S1 illustrates the method used to prepare pentagon and heptagon-shaped samples using a scanning electron microscope and $Ga^+$ focused ion beam (FIB-SEM) machine. A lift-out procedure is first performed to extract a piece of sample from a bulk crystal [1]. A $\sim 2 \times 2 \times 10$ µm$^3$ layer of carbon is deposited locally to protect the surface of the sample (a,b) using electron beam-induced and ion beam-induced deposition (EBID and IBID). The area around the deposit is then milled (c) to free a piece of sample of approximately $10 \times 2 \times 5$ µm$^3$. The piece of sample is extracted using a micro-probe (d) and attached to a 3 mm copper grid (e) using again IBID of carbon. After the probe is released, the surfaces of the sample can be polished and flattened, which is recommended for the next step. The sample is then flipped 90° so that it is perpendicular to the ion beam for patterning the desired geometry (g), in this case a pentagon. A 30 kV and 80 pA $Ga^+$ ion beam was used for this step. The milling duration should not be too long so that it does not completely pass through the sample. However, it should be long enough to reach approximately 500 nm deep so that surface damage layers can be removed later. The milling duration can be calibrated by sacrificing a small region of the sample to check how much time is needed to go through. After patterning, the milled regions are filled with carbon using IBID and the same pattern as for milling (h). The sample is then flipped back to its original orientation to thin it down to a thickness of approximately 150 nm. The backside of the sample (the patterned side is considered as the front side) is first milled until the pattern becomes visible (j,k) in SEM. Finally, the front side can then be polished to remove surface damages created during the previous steps (l). Fig. S1(m,n) shows, for example, side-view and top-view SEM images of a TEM lamella patterned with pentagons of different sizes.

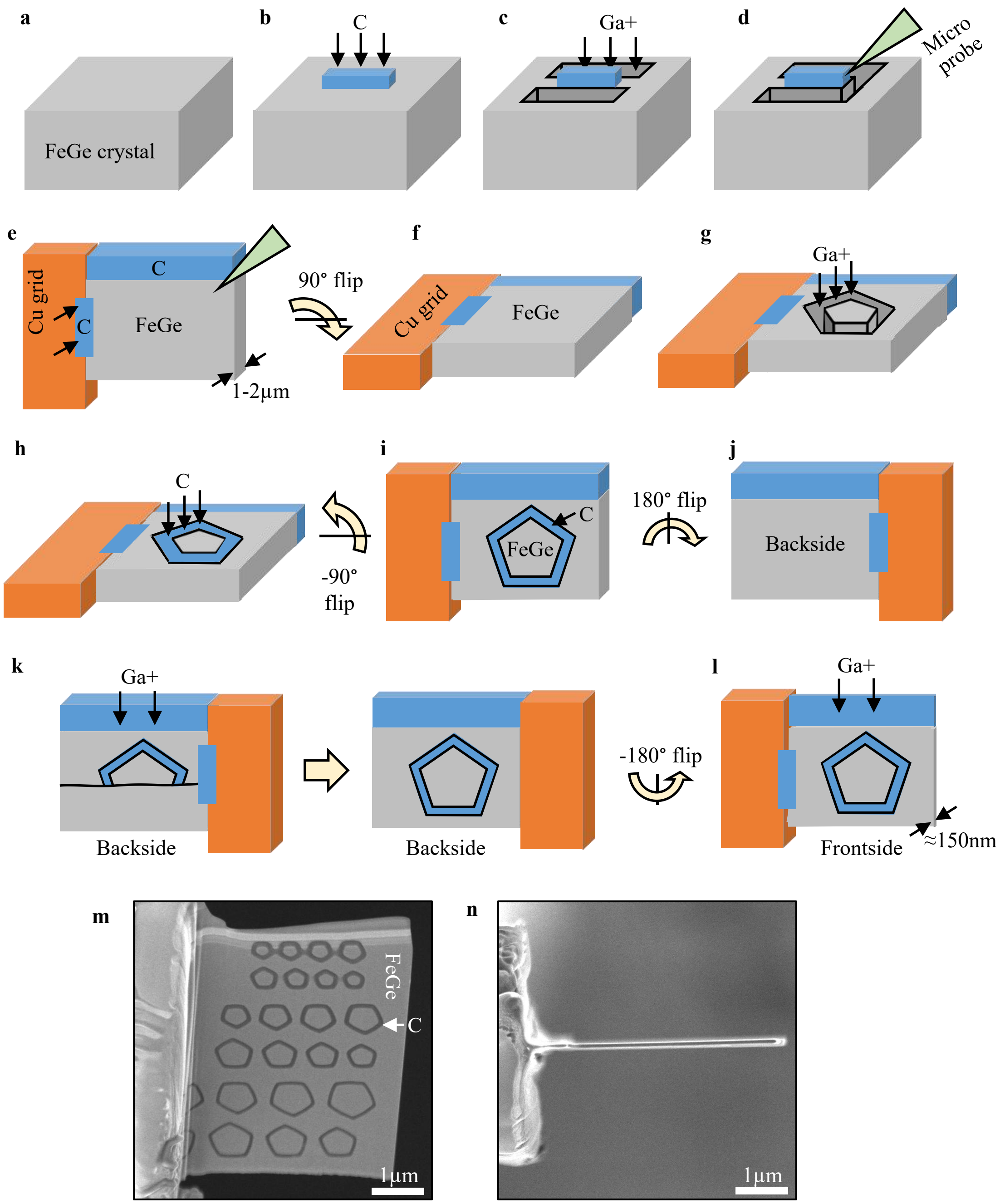

Figure S1: **FIB preparation of a TEM lamella with pentagon-shaped patterns.** (a) Bulk FeGe crystal. (b) Deposit a capping layer of carbon using EBID and IBID. (c) Mill areas around the capping layer. (d) Attach a micro-probe to the sample and free the remaining parts. (e) Attach the sample to a copper grid. (f) Flip the sample 90° around a horizontal axis. (g) Pattern the surface of the sample. (h) Deposit carbon in the patterned areas. (l) Flip the sample back to 0°. (j) Rotate the sample 180° around a vertical axis to expose the backside. (k) Mill the backside until the pattern becomes visible. (l) Rotate the sample back to 0° and thin it down to electron transparency. (m,n) Side-view and top-view SEM images of an example of lamella patterned with pentagons of different sizes.

## Supplementary Information 2: Applied field series of pentagon- and heptagon-shaped samples

Figure S2(a,b) shows Fresnel images of a pentagon- and a heptagon-shaped samples. The images were obtained at 220 K and in the presence of different external magnetic fields indicated in the figure. It shows the evolution of the magnetic domain structure from the helical state at low fields to the skyrmion lattice state at approximately 200 mT and then a saturated monodomain state at approximately 400 mT.

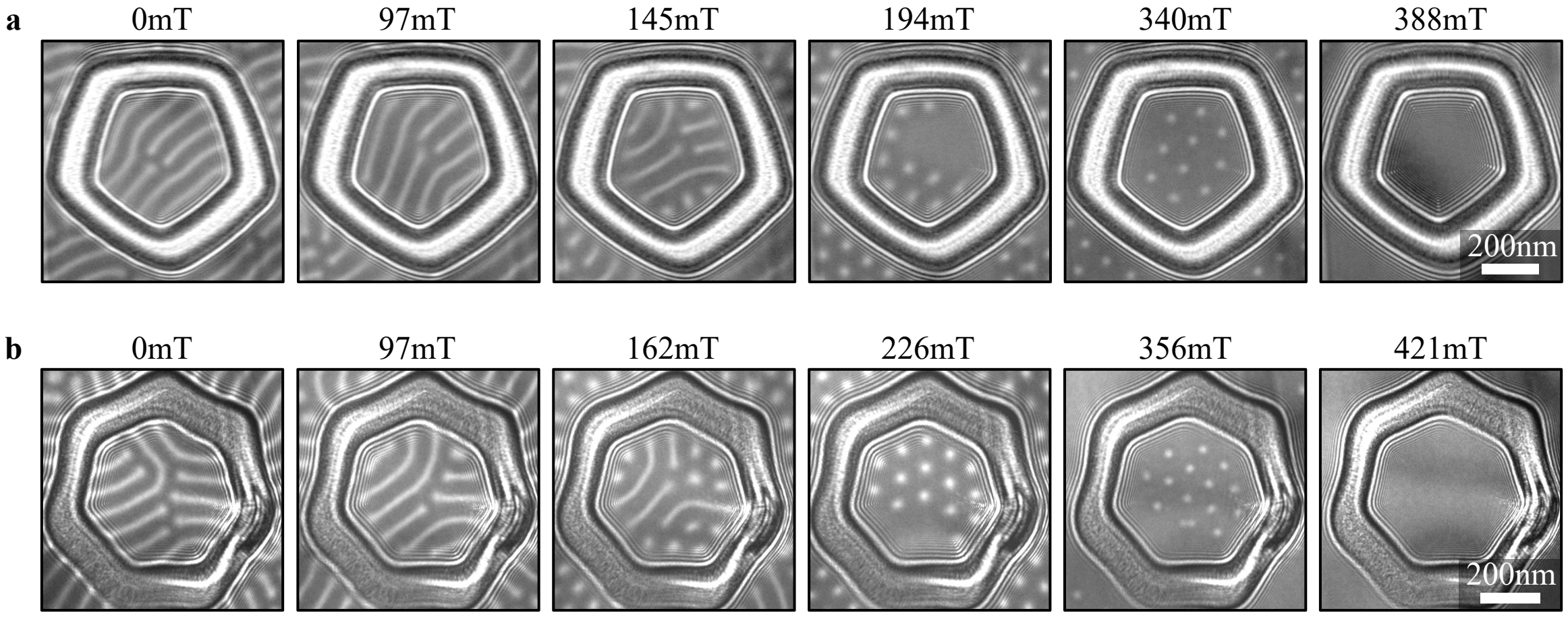

Figure S2: **LTEM applied field series**. (a) Fresnel image of a pentagon-shaped sample recorded at 220 K with a defocus of 500 µm and in the presence of different external magnetic fields indicated in the figure. (b) Same description in the case of a heptagon-shaped sample.

## Supplementary Information 3: Skyrmion lattices in a heptagon-shaped sample

Figure S3(a) shows a TEM image of a heptagon-shaped sample with a side length of $l = 240$ nm. As shown schematically in Fig. S3(b), the total number of skyrmions in a 7-fold disclination state is

$$S = 1 + \sum_{n=1}^{N} 7n\,, \tag{S1}$$

with in this case $N = 2$ skyrmion rings and $S = 22$. Figure S3(c) shows Fresnel images of lattices with different skyrmion numbers obtained at 220 K and with an external magnetic field of 240 mT. In the 7-fold disclination with $S = 22$, the central skyrmion $n = 0$ has seven neighbors and the skyrmions of the inner ring $n = 1$ have six neighbors as shown schematically in Fig. S3(d). Interestingly, when $S = 21$, two skyrmions are removed from the outer ring $n = 2$ and one is added at the center $n = 0$. Therefore, the lattice shows two interconnected pentagons at the center. When $S = 20$, one skyrmion is removed from the center with respect to $S = 21$, forming again a heptagon at the center with two pentagons on the sides. On the other hand, when $S = 23$, one skyrmion is added to $n = 0$ with respect to $S = 22$ forming two pentagons at the center and two heptagons on the sides. These observations are in agreement with micromagnetic simulations shown in Fig. S3(e).

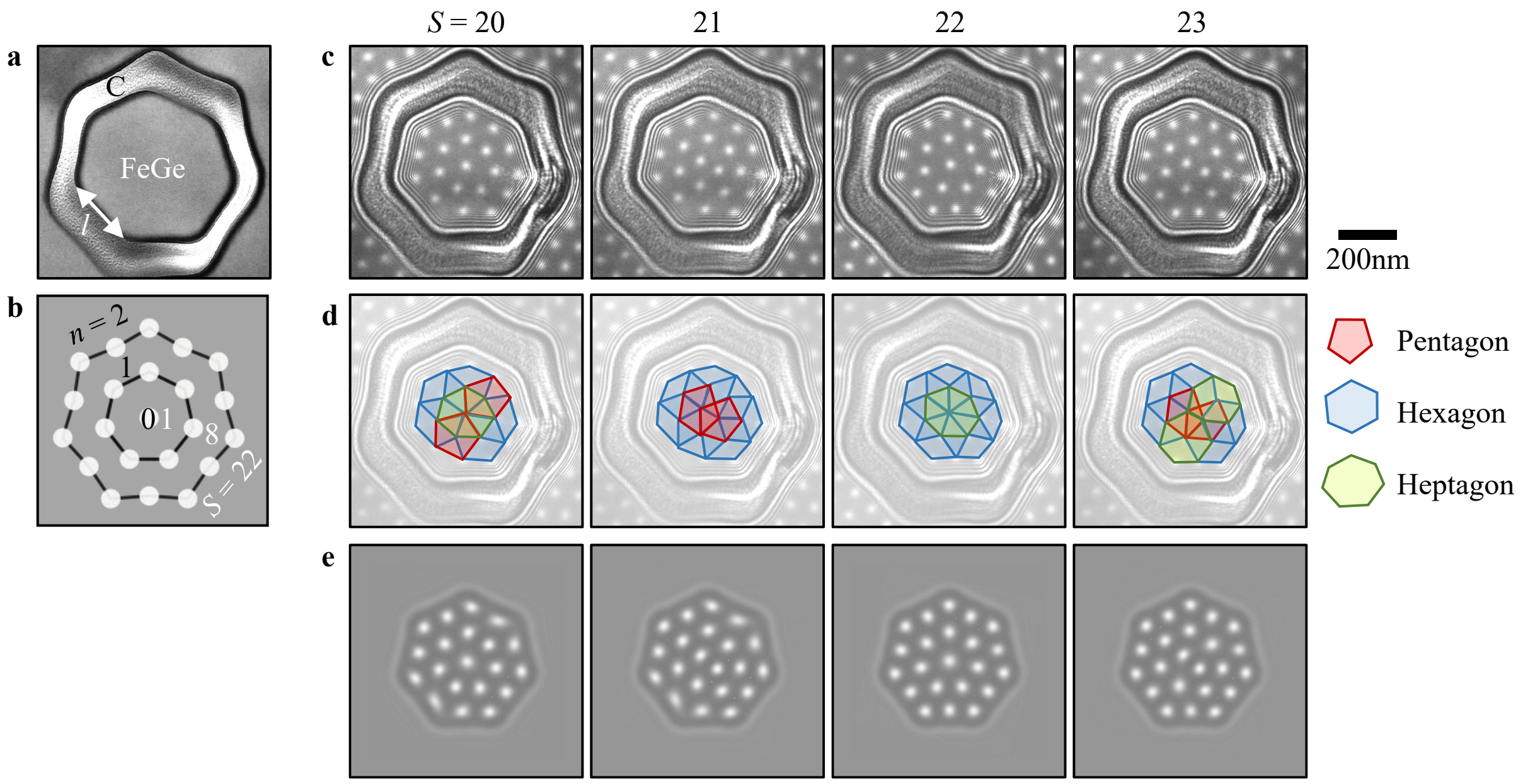

Figure S3: **LTEM of skyrmion lattices in a heptagon-shaped sample.** (a) In-focus TEM image of a heptagon-shaped sample with a side length of $a = 240$ nm. (b) Schematic showing a 7-fold disclination with 22 skyrmions. (c) Experimental Fresnel images obtained at 220 K with a defocus of 500 µm and in the presence of an external magnetic field of 240 mT. The images contain lattices with different skyrmion numbers indicated in the figure. (d) Previous images overlaid with pentagons, hexagons and heptagons. (e) Fresnel images calculated from micromagnetic simulations.

# Supplementary Information 4: Field evolution of a 5-fold disclination

Figure S4(a) shows an applied field series of *magnetic* phase images $\phi_{mag}$ of a 5-fold disclination in a pentagon-shaped sample. Figure S4(b,c) shows the corresponding magnitude and direction of the magnetic induction field $\boldsymbol{B}_{\perp}$. At low field, for instance 176 mT, the central skyrmion shows a pentagon shape and the surrounding skyrmions show a hexagonal shape that can be elongated along the circular direction around the disclination. When increasing the field, the skyrmions appear progressively smaller and more round-shaped.

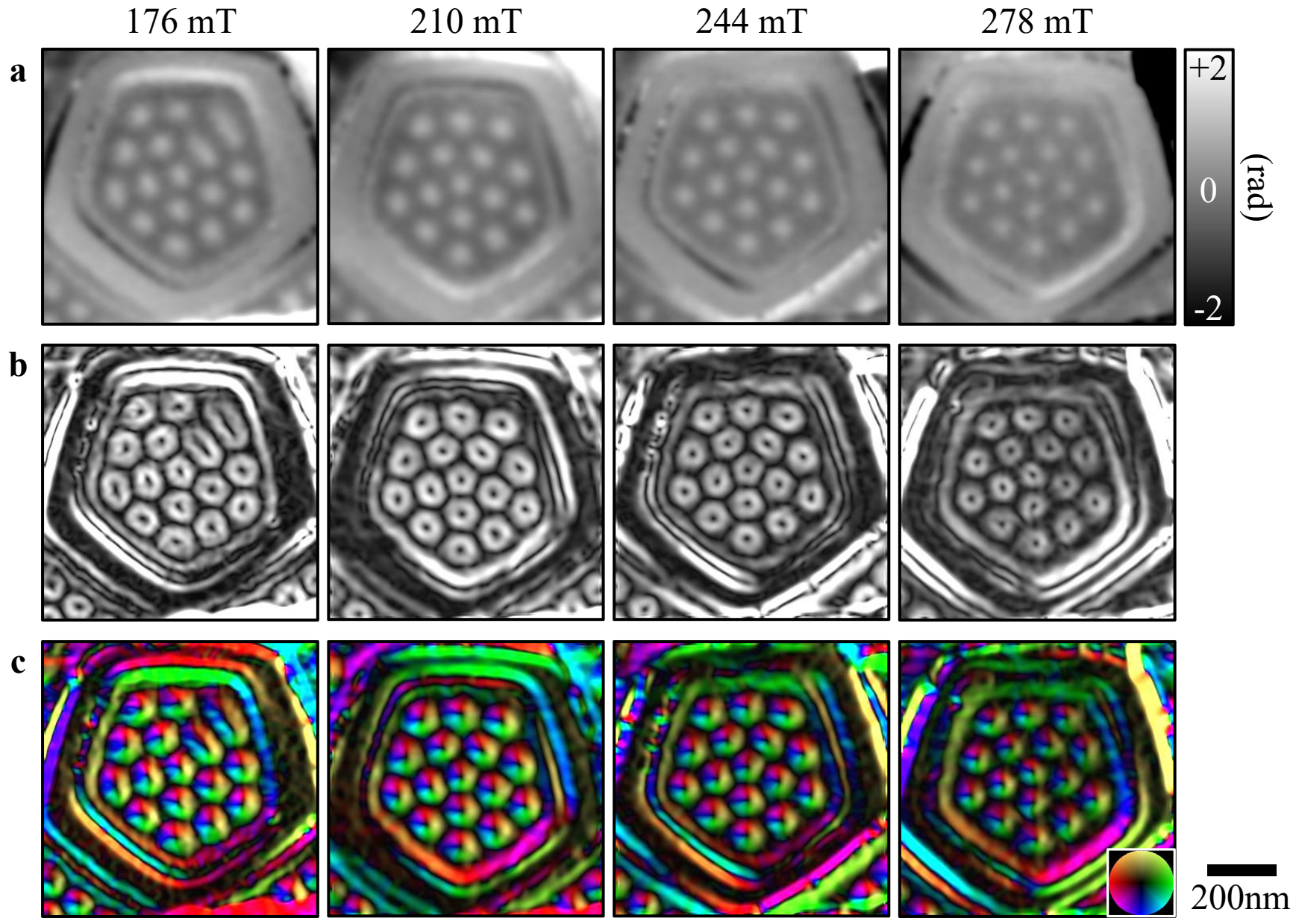

Figure S4: **Electron holography of a 5-fold disclination.** (a) Magnetic phase images $\phi_{mag}$ of a pentagon-shaped sample obtained at 220 K and with different external magnetic fields indicated in the figure. (b) Magnitude of the in-plane magnetic induction field $|\boldsymbol{B}_{\perp}|$. (c) Corresponding color-coded maps showing the direction of $\boldsymbol{B}_{\perp}$.

# Supplementary Information 5: Magnetic and structural analysis of a 7-fold disclination

Figure S5(a) shows an off-axis electron hologram of the heptagon-shaped sample that hosts a 7-fold disclination. Fig. S5(b-e) shows the reconstructed *magnetic* phase image $\phi_{mag}$, the magnitude and direction of the in-plane magnetic induction field $\boldsymbol{B}_{\perp}$, and profiles extracted from the phase image as indicated by the solid and dashed lines. It can be observed in (c,d) that the central skyrmion is slightly larger than the other skyrmions around it. The profiles in (e) show that the inter-skyrmion distance along the radial direction (solid red profile) is larger that the inter-skyrmion distance along the edge direction (dashed blue profile). The addition of a $2\pi/7$ wedge induces a compression of the seven wedges along the circular direction around the disclination. Figure S5(f-j) shows corresponding micromagnetic simulations, in good agreement with the experiment.

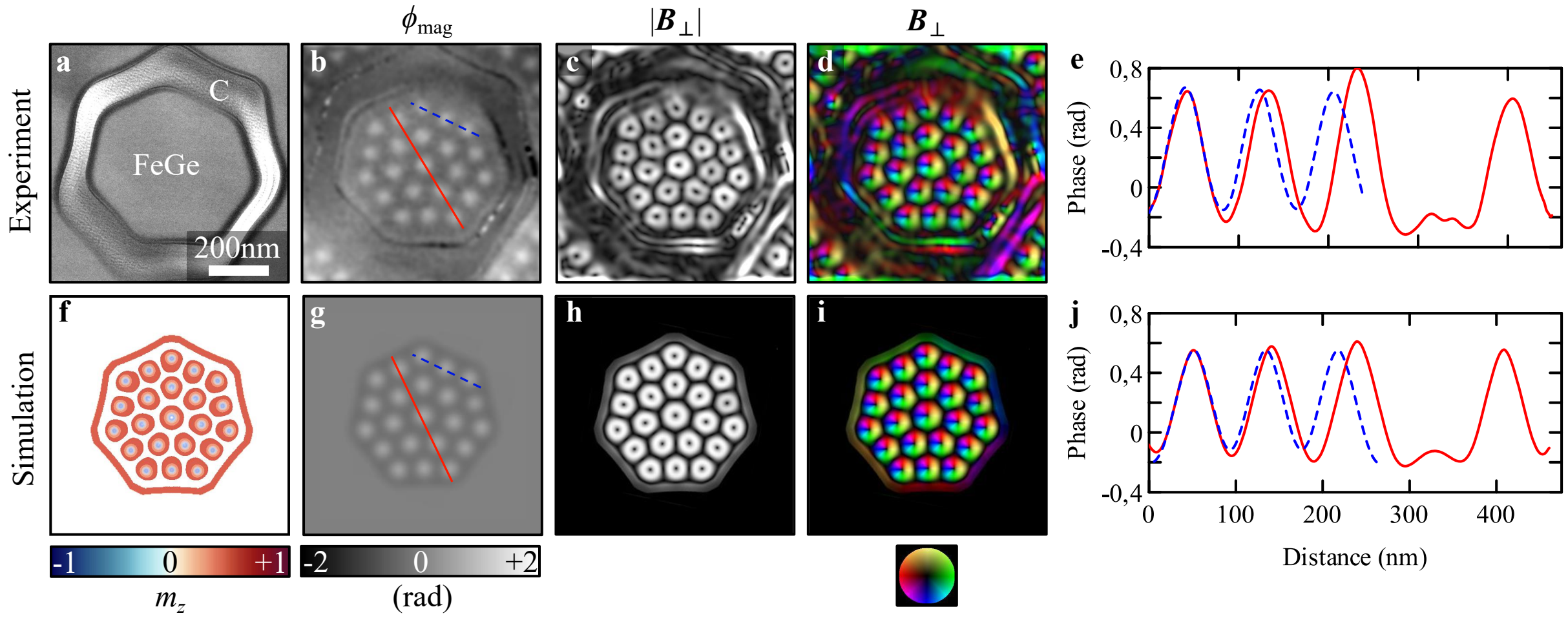

Figure S5: **Electron holography of a 7-fold disclination.** (a) Electron hologram of a pentagon-shaped sample recorded with an external magnetic field of 240 mT. (b) Magnetic phase image $\phi_{mag}$ obtained using off-axis electron holography. (c) Magnitude of the in-plane magnetic induction field $|\boldsymbol{B}_{\perp}|$. (d) Corresponding color-coded map showing the direction of $\boldsymbol{B}_{\perp}$. (e) Phase profiles extracted from (b) along the solid and dashed lines. (f) Isosurface representation of the normalized out-plane component $m_z$ of a magnetization model calculated using micromagnetic simulation. (g-j) Corresponding phase image, magnetic field maps and profiles.

To investigate the structure of the disclination, the coordinates of the skyrmions in the phase image shown in Fig. S6(a) were determined by calculating their center of mass. The local orientational parameter of the 7 hexagons was then calculated, as explained in the main article. Figures S6(b) and (c) represent respectively the orientation $\theta$ of the hexagons and their degree of hexagonality $|\Psi_6|$. The orientation $\theta$ changes by approximately 9° and in average the hexagonality is $\langle|\Psi_6|\rangle = 0.85$. For comparison, the white dots in Fig. S6(d) show a 7-fold disclination model created from a hexagonal lattice (black dots) using linear elastic theory. Figure S6(e) and (f) show the corresponding rotation and hexagonality maps calculated from the positions in (d). The rotation changes by steps of $\Delta\theta \approx 9°$ along the circular direction and the hexagonality is $|\Psi_6| = 0.91$, which is close to the experimental values.

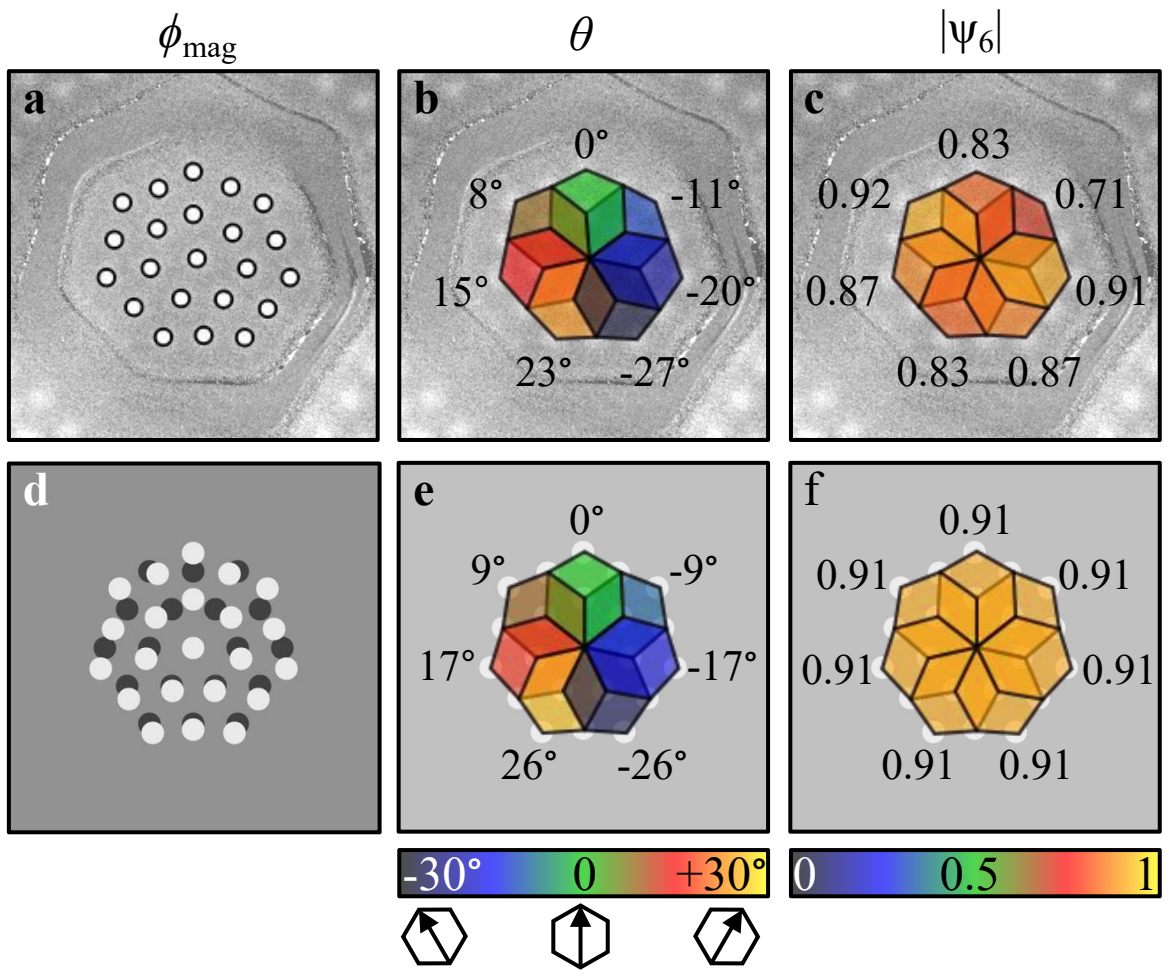

Figure S6: **Structural analysis and linear elastic theory.** (a) Position of the skyrmions in the magnetic phase image. (b) Orientation $\theta$ and (c) hexagonality $|\Psi_6|$ maps calculated form the positions in (a). (d) 7-fold disclination model calculated using linear elastic theory. (e) Corresponding orientation and (f) hexagonality maps.

# Supplementary Information 6: Disclinations in pentagons and heptagons of different sizes

Figure S7(a-c) shows experimental LTEM images obtained in pentagons of increasing sizes that contain disclinations with (a) $N = 1$, (b) 2 and (c) 3 skyrmion rings. Figure S7(d-f) shows heptagons that contain the corresponding 7-fold disclinations. In practice, the probability to obtain large disclinations by cycling the external field is low and a large number of attempts were needed to obtain the disclinations with $N = 3$ rings.

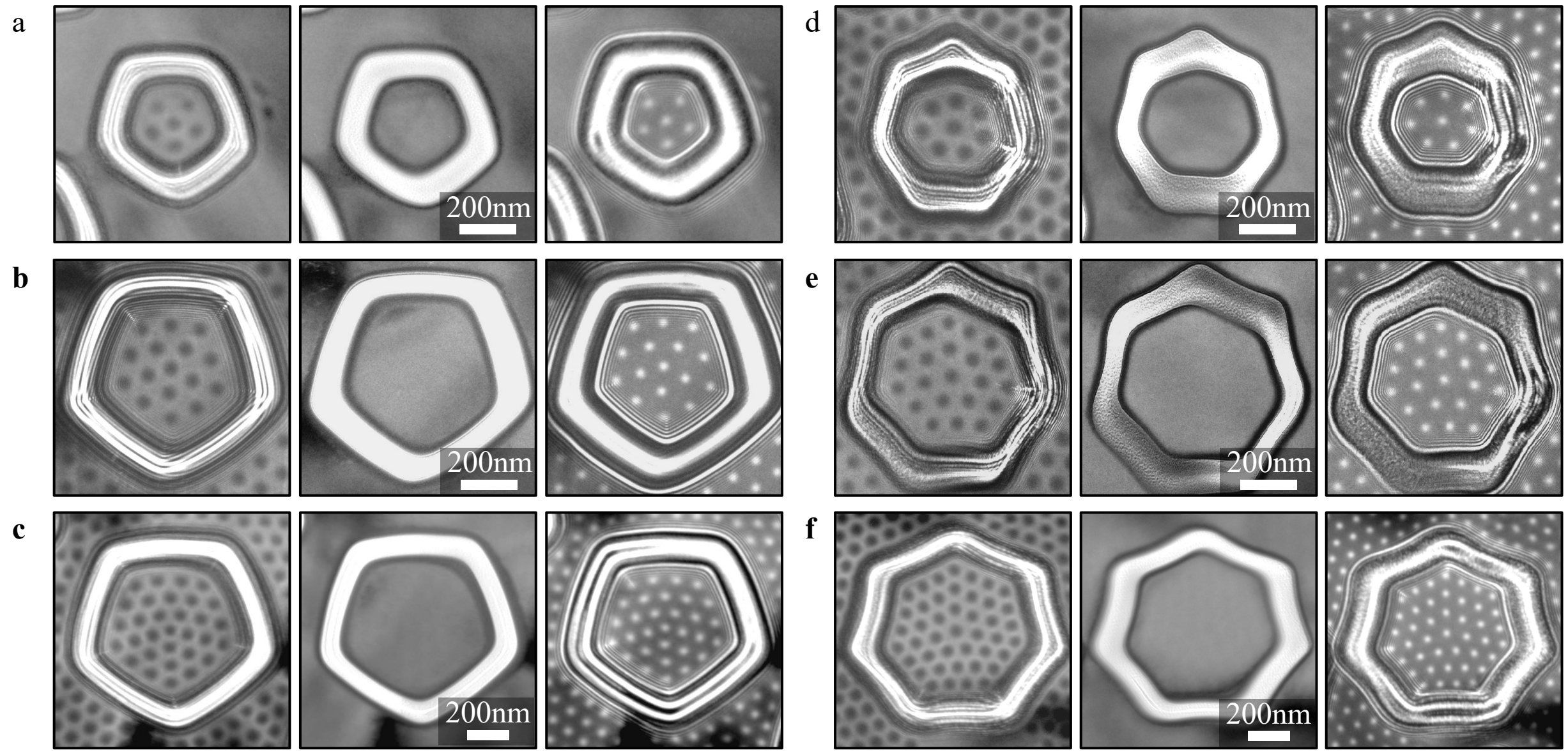

Figure S7: **LTEM imaging of skyrmion lattices in pentagon- and heptagon-shaped samples of different sizes.** (a-c) Pentagons that host 5-fold disclinations with $N = 1$ to 3 skyrmions rings observed at −500, 0, and +500 µm defocus. (d-f) Same description with heptagons that host 7-fold disclinations.

# Supplementary Information 7: Skyrmions mobility around a disclination

It was observed in the main article that when $S = 17$ skyrmions in a pentagon, a 5-7 dislocation is created. The 7-coordinated skyrmion of the dislocation can be located in any of the five different corners of the pentagon leading to five geometrically equivalent configurations. It is possible to switch between the different configurations by tilting the sample with respect to the applied out-of-plane magnetic field and so adding a small in-plane component to it. Figure S8(a) shows a similar situation but with a larger pentagon that contains three rings with a total of 32 skyrmions. There are two 5-7 dislocations indicated by arrows in Fig. S8(b). When the sample is tilted between -3° and +3°, the position of the dislocations changes from the edges to the corners of the pentagon. At an intermediate tilt of 0°, the outer skyrmion ring at $n = 3$ appears diffuse along the circular direction because the lattice wobbles between the two configurations. On the other hand, the skyrmions of the inner rings stay sharp. In Fig. S8(c), the position of the skyrmions observed at +3° and -3° is indicated by blue and red circles. It can be observed that switching between the two configurations involves a displacement along the circular direction of most of the skyrmions of the outer ring, which can then explain its diffuse appearance at 0°.

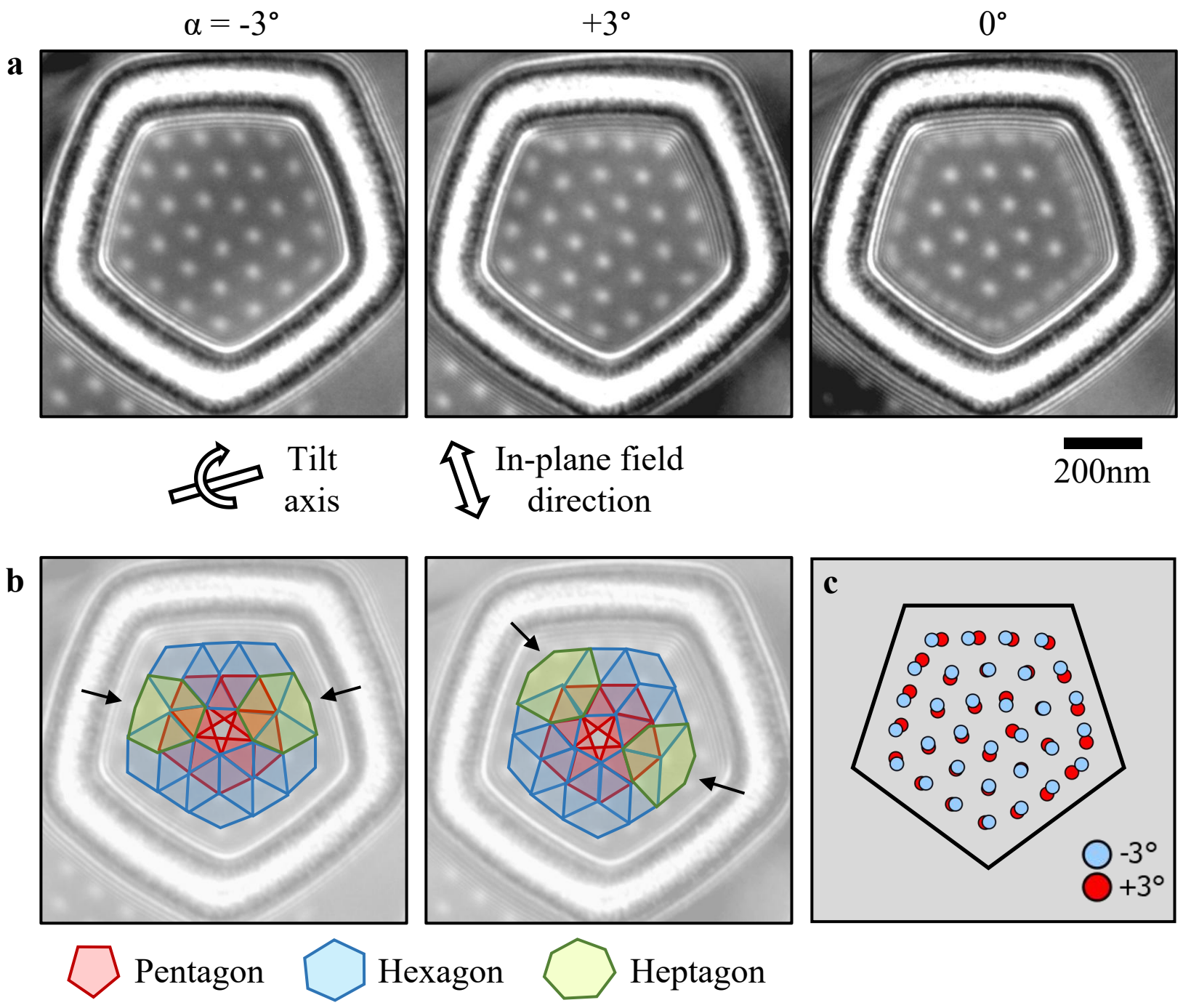

Figure S8: **Skyrmions mobility around a disclination.** (a) Fresnel images of a skyrmion lattice with $S = 32$ skyrmion in a pentagon-shaped sample obtained at 220 K with a defocus of 500 µm, in the presence of an external magnetic field of 230 mT and at different tilt angles $\alpha = -3°$, $+3°$ and 0°. (b) Images obtained at -3° and +3° overlaid with pentagons, hexagons and heptagons. (c) Positions of the skyrmions in the two tilted images represented by blue and red dots.